
\tolerance=1500
\baselineskip=15pt
\magnification=\magstep1
\raggedbottom
\centerline{\bf INFLATING LORENTZIAN WORMHOLES}
\vskip 0.3in
\centerline{Thomas A. Roman}
\centerline{Physics and Earth Sciences Department}
\centerline{Central Connecticut State University, New Britain,
Connecticut 06050}
\vskip 0.3in
\centerline{\bf Abstract}
\vskip 0.2in
        It has been speculated that Lorentzian wormholes of
the Morris-Thorne type might be allowed by the laws of physics
at submicroscopic, e.g. Planck, scales and that a sufficiently
advanced civilization might be able to enlarge them to classical
size. The purpose of this paper is to explore the possibility
that inflation might provide a natural mechanism for the
enlargement of such wormholes to macroscopic size. A new classical
metric is presented for a Lorentzian wormhole which is imbedded in
a flat deSitter space. It is shown that the throat and the proper
length of the wormhole inflate. The resulting properties and
stress-energy tensor associated with this metric are discussed.
\eject

\centerline{\bf I. INTRODUCTION}
\vskip 0.2in
      There has been much interest recently in the Lorentzian
signature, traversable wormholes conjectured by Morris and Thorne
(MT) [1,2]. These wormholes have no horizons and thus allow two-way
passage through them. As a result, violations of all known energy
conditions, including the weak (WEC) [3] and averaged weak (AWEC)
[4] energy conditions, must unavoidably occur at the throat of the
wormhole. Another disturbing (or intriguing, depending on one's
point of view) property of these wormholes is the possibility of
transforming them into time machines for backward time travel
[5,6] and thereby, perhaps, for causality violation. Whether such
wormholes are actually allowed by the laws of physics is currently
unknown. However, recent research by Hawking [7] and others [8]
indicates that it is very likely that nature employs a ``Chronology
Protection Agency'' which prevents the formation of closed timelike
curves. The method of enforcement appears to be the divergences in
vacuum expectation values of the stress-energy tensor of test
fields which accompany the advent of any self-intersecting null
geodesics. It appears that this behavior is generic with the
formation of closed timelike curves [7,8]. At this point it is
not clear whether these results imply that traversable wormholes
cannot exist at all or that nature just does not permit their
conversion into time machines.

      It has been known for some time that quantum field theory
allows local violations of the WEC [9] in the
form of locally negative energy densities and fluxes, the most
notable example being the Casimir Effect [10]. A major unresolved
issue is whether quantum field theory permits the macroscopic
effects of negative energy required to maintain traversable
wormholes  against collapse. Wald and Yurtsever [11] have recently
shown that the AWEC condition holds for massless
scalar fields in a wide range of spacetimes, but that it apparently
does not hold in an arbitrary curved four-dimensional spacetime. It
is possible that although violations of the WEC (or AWEC) might be
allowed, the magnitude and duration of
these violations may be limited by uncertainty principle-type
inequalities which could render gross macroscopic effects of
negative energy unobservable. This appears to be the case for
negative energy fluxes due to quantum coherence effects in
flat spacetime [12]. Such quantum inequalities also appear to
prevent the unambiguous observation of violations of cosmic
censorship in the attempt to produce a naked singularity
from an extreme Reissner-Nordstrom black hole, in both two and
four dimensions [13]. Quantum inequalities also constrain the
magnitude and duration of the negative energy flux seen by
an observer freely falling into an evaporating two-dimensional
Schwarzschild black hole [14].

      Several equally important, though much less explored,
questions are: A) Do the laws of physics permit the topology
change required to create the wormhole in the first place?
In classical general relativity, such topology change
must be accompanied by the creation of closed timelike curves
[7,15]. Also, at least some topology change issues may be related
to energy conditions [16]. B) Do the laws of physics permit
submicroscopic Lorentzian wormholes (e.g. on the Planck scale
[17].)? It may be that wormhole formation, although possibly
prohibited on the classical level, might be allowed
quantum-mechanically. If so, then: C) Are there processes, either
natural or artificial, which could lead to their enlargement to
classical size? The present paper will attempt to address one
aspect of the last question.

      MT suggest that, ``One can imagine an
advanced civilization pulling a wormhole out of the quantum foam
and enlarging it to classical size.'' This would seem to be, at
best, wishful thinking. However, consider the following scenario.
Suppose that a submicroscopic MT-type wormhole could form in the
very early universe via, say, a quantum fluctuation (the nature of
which we will leave conveniently vague). Is it possible that
subsequent inflation of the universe, if it occurs, could enlarge
the wormhole to classical size? Or perhaps it might be possible
to artificially enlarge a tiny wormhole by imbedding it in a false
vacuum bubble, as in the ``creation of a universe in the
laboratory'' scenario [18].
       The inflation of quantum fluctuations
of a scalar field has previously been invoked as a mechanism for
providing the seeds of galaxy formation [19]. Basu et al.[20]
have examined the nucleation and evolution of topological defects
during inflation. Mallett [21]
has modeled the effects of inflation on the evaporation of a black
hole using a Vaidya metric imbedded in a deSitter background. His
results suggest that inflation depresses the rate of black hole
evaporation. Sato et al. [22] have studied the formation of a
Schwarzschild-deSitter wormhole in an inflationary universe.
More recently, Kim [23] has constructed a traversable wormhole
solution by gluing together two Schwarzschild-deSitter metrics
across a $\delta$-function boundary layer, following the methods of
Visser [24]. Hochberg [25] has used a similar technique to
construct Lorentzian wormhole solutions in higher derivative
gravity theories. Hochberg and Kephart [26] have argued that
gravitational squeezing of the vacuum might provide a natural
mechanism for the production of the negative energy densities
required for wormhole support. However, recent work of Kuo and Ford
[27] indicates that many states of quantized fields which involve
negative energy densities are accompanied by large fluctuations in
the expectation value of the stress-energy tensor. For such states
the semiclassical theory of gravity may not be a good
approximation. The states they examined included squeezed states
and the Casimir vacuum state.

       The outline of the present paper is as follows. In Sec.II,
a new class of metrics is presented which represents a Lorentzian
wormhole imbedded in a deSitter inflationary background. The
imbedding is quite ``natural'' in that it does not involve
``thin-shells'' or $\delta$-function ``transition layers''. The
stress-energy tensor of the false vacuum for deSitter space barely
satisfies the weak energy condition, since the energy density is
exactly equal to minus the pressure. So these models couple
``exotic'' (i.e., energy-condition violating) to ``near-exotic''
matter, in the terminology of MT. In the limit of vanishing
cosmological constant, the metric reduces to the static MT
traversable wormhole. It is demonstrated that both the throat and
the proper length of the wormhole inflate. The resulting stress-
energy tensor is constructed by plugging the metric into the
Einstein equations. (Although it is possible that such a metric
might represent a wormhole which was ``caught'' in an inflationary
transition, to definitively show this one would need to solve
the opposite problem. That is, one would have to come up with
a physically plausible stress-energy tensor and solve the Einstein
equations to find the metric, and then show that the resulting solution
had the desired wormhole characteristics. This is a much more
difficult problem than the one treated here). The properties of our
metrics are discussed in Sec.III. We use the same metric and curvature
conventions as MT [1], and we work in units where $G=c=1$.
\eject
\centerline{\bf II. A MORRIS-THORNE WORMHOLE IN AN INFLATING
BACKGROUND}
\vskip 0.5in
\centerline{A. A Review of Static Morris-Thorne Wormholes}
\vskip 0.2in

       To make this paper relatively self-contained, we will review
the results of MT [1]. The metric for a general
MT traversable wormhole is given by
 $$ds^2=-e^{2\Phi(r)}dt^2+{{dr^2}\over{(1-b(r)/r)}}
           +r^2({d\theta}^2+ sin^2\theta\,{d\phi}^2)\,,
                                                \eqno(2.1)$$
where the two adjustable functions $b(r)$ and $\Phi(r)$ are
referred to as the ``shape function'' and the ``redshift
function'', respectively. The shape function $b(r)$ controls the
shape of the wormhole as viewed, for example, in an embedding
diagram. The metric Eq. (2.1) is spherically symmetric and static.
The geometric significance of the radial coordinate $r$ is that the
circumference of a circle centered on the throat of the wormhole is
given by $2\pi r$. The coordinate $r$
is nonmonotonic in that it decreases from $+\infty$ to a minimum
value $b_o$, representing the location of the throat of the
wormhole, and then it increases from $b_o$ to $+\infty$. This
behavior of the radial coordinate reflects the fact that the
wormhole connects two separate external ``universes'' (or two
regions of the same universe). At the throat, defined by $r=b=b_o$,
there is a coordinate singularity where the metric coefficient
$g_{rr}$ becomes divergent, but the radial
proper distance
      $$l(r)=\pm\,\int_{b_o}^r{{dr}\over{(1-b(r)/r)^{1/2}}}\,
                                                 \eqno(2.2)$$
must be required to be finite everywhere. At the throat $l=0$,
while $l<0$ on the ``left'' side of the throat and $l>0$ on the
``right'' side. For the wormhole to be traversable it must have
no horizons, which implies that $g_{tt}=-e^{2\Phi(r)}$ must never
be allowed to vanish. This condition in turn imposes the constraint
that $\Phi(r)$ must be finite everywhere.

    To construct an embedding diagram [1,28] of the wormhole one
considers the geometry of a $t=const.$ slice. Using the spherical
symmetry, we can set $\theta={\pi}/2$ (an ``equatorial'' slice).
The metric on the resulting two-surface is
    $$ds^2={{dr^2}\over{(1-b(r)/r)}} + r^2{d\phi}^2\,.
                                                  \eqno(2.3)$$
The three-dimensional Euclidean embedding space metric can be
written as
    $$ds^2=dz^2+dr^2+r^2{d\phi}^2\,.  \eqno(2.4)$$
Since the embedded surface is axially symmetric, it can be
described by $z=z(r)$, sometimes called the ``lift function''
(see [1,28]). The metric on the embedded surface can then be
expressed as
     $$ds^2=\left[1+\left({{dz}\over{dr}}\right)^2\right]\,dr^2
            +r^2{d\phi}^2\,.             \eqno(2.5)$$
Equation (2.5) will be the same as Eq. (2.4) if we identify the
$r$, $\phi$ coordinates of the embedding space with those of the
wormhole spacetime, and also require:
     $${{dz}\over{dr}}=\pm\left({{r}\over{b(r)}}-1\right)^{-1/2}\,.
                                         \eqno(2.6)$$
A graph of $z(r)$ yields the characteristic wormhole pictures found
in [1,28]. For the space to be asymptotically flat far from the
throat, MT require that ${dz}/{dr}\rightarrow\,0$ as
$l\rightarrow\pm\infty$, i.e., $b/r\rightarrow\,0$ as
$l\rightarrow\pm\infty$. In order for this condition to be
satisfied, the wormhole must flare outward near the throat, i.e.,
          $${{d^2r(z)}\over{dz^2}}>0\,,  \eqno(2.7)$$
at or near the throat. Therefore
          $${{d\,^2r(z)}\over{dz^2}}={{{b-b'r}\over{2b^2}}}>0\,,
                                        \eqno(2.8)$$
at or near the throat, $r=b=b_o$, where the prime denotes
differentiation with respect to $r$.

     MT define an ``exoticity function'':
          $$\zeta\equiv{{\tau-\rho}\over{|\rho|}}
               ={{b/r-b'-2(r-b){\Phi}'}\over{|b'|}}\,,
                                           \eqno(2.9)$$
where $\rho$ and $\tau$ are the energy density and radial tension,
respectively, as measured by static observers in an orthonormal
frame. MT show that Eq. (2.9) can be written as
    $$\zeta={{2b^2}\over{r |b'|}}\left({{d^2r(z)}\over{dz^2}}
         \right)-{{2(r-b){\Phi}'}\over{|b'|}}\,, \eqno(2.10)$$
and argue (see Sec.III.F2 of MT) that Eq. (2.10) reduces to
    $$\zeta_o={{\tau_o-\rho_o}\over{|\rho_o|}}>0\,, \eqno(2.11)$$
at or near $r=b=b_o$.

       The general strategy is then to choose $\Phi(r)$ and $b(r)$
to get a ``nice'' wormhole, and to compute the resulting
stress-energy tensor components by plugging $\Phi$, $b$ into the
Einstein equations. One can show quite generally [1,5] that the
resulting stress-energy tensor must violate all known energy
conditions, including both the WEC and AWEC. It
is known that quantum fields can
violate the WEC [9]. Whether or not the laws of quantum field
theory permit violations of AWEC large enough to support a
macroscopic (or microscopic, for that matter) traversable wormhole
is presently unknown [11].

       One class of particularly simple solutions considered by MT
are the so-called ``zero-tidal-force'' solutions, corresponding to
the choice $b=b(r)$, $\Phi(r)=0$. The choice of $\Phi=0$ yields
zero tidal force as seen by stationary observers. We write the
metric for later reference as
       $$ds^2=-dt^2+{{dr^2}\over{(1-b(r)/r)}}
           +r^2({d\theta}^2+ sin^2\theta\,{d\phi}^2)\,.
                                                     \eqno(2.12)$$
The energy density $\rho(r)$, radial tension per unit area
$\tau(r)$, and lateral pressure $p(r)$ for this class of wormholes
as seen by static observers in an orthonormal frame are given by
       $$T_{\hat t \hat t}=\rho(r)={{b'(r)}\over{8\pi r^2}}
                                                      \eqno(2.13)$$
       $$-T_{\hat r \hat r}=\tau(r)={{b(r)}\over{8\pi r^3}}
                                                     \eqno(2.14)$$
       $$T_{\hat \theta \hat \theta}=T_{\hat \phi \hat \phi}
         =p(r)={{b(r)-b'r}\over{16\pi r^3}}\,.       \eqno(2.15)$$

       Two examples of this class of wormholes are the following.
The first is given by:
         $$b(r)={{b_o}^2\over r} \,,\, \Phi(r)=0\,.   \eqno(2.16)$$
This corresponds to
         $$z(r)=b_o cosh^{-1}\left({r\over{b_o}}\right)\,,
                                                    \eqno(2.17)$$
which has the shape of a catenary, i.e.,
         $${{dz}\over{dr}}={{b_o}\over{\sqrt{r^2-{b_o}^2}}}\,.
                                                    \eqno(2.18)$$
The wormhole material is everywhere exotic, i.e., $\zeta>0$
everywhere. It extends outward from the throat, with $\rho,\,\tau$,
and $p$ asymptoting to zero as $l=\pm\infty$.

        The second example corresponds to the confinement of the
exotic matter to an arbitrarily small region around the throat. MT
call this an ``absurdly benign'' wormhole. It is given by the
choice:
    $$b(r)=\cases{b_o[1-{(r-b_o)}/{a_o}]^2\,\,,\, \Phi(r)=0 &
for $b_o\leq r\leq b_o+a_o$,\cr
&\cr
b=\Phi=0& for $r\geq b_o+a_o$.\cr}               \eqno(2.19)$$
For $b_o<r<b_o+a_o$,
     $$\rho(r)=\left[\left(-{b_o}/{a_o}\right)/{(4\pi
r^2)}\right]\left[1-{(r-b_o)}/{a_o}\right]<0   \eqno(2.20)$$
     $$\tau(r)={b_o \left[1-{(r-b_o)}/{a_o}\right]^2}/
{(8\pi r^3)}                                       \eqno(2.21)$$
     $$p(r)={1\over2}(\tau-\rho)\,.                \eqno(2.22)$$
For $r\geq b_o+a_o$, the spacetime is Minkowski, and
$\rho=\tau=p=0$.
\vskip 0.5in

\centerline{B. The $\Phi(r)\neq 0$ Inflating Wormholes}
\vskip 0.2in
        A simple generalization of the original MT wormhole
metrics, characterized by Eq. (2.1), to a time-dependent
inflationary background is:
$$ds^2=-e^{2\Phi(r)}dt^2+e^{2\chi t}\left[{{dr^2}\over{(1-b(r)/r)}}
   +r^2({d\theta}^2+ sin^2\theta\,{d\phi}^2)\right]\,,
                                                   \eqno(2.23)$$
Here we have simply multiplied the spatial part of the metric
Eq. (2.1), by a deSitter scale factor $e^{2\chi t}$, where
$\chi=\sqrt{\Lambda/3}$ and $\Lambda$ is the
cosmological constant [29].
The coordinates $r,\theta,\phi$ are chosen to have the same
geometrical interpretation as before. In particular, circles of
constant $r$ are centered on the throat of the wormhole. Our
coordinate system is chosen to be
``co-moving'' with the wormhole geometry in the sense that the
throat of the wormhole is always located at $r=b=b_o$ for all $t$.
(Of course, this does
not mean that two points at different (constant) values of
$r,\theta,\phi$ have constant {\it proper} distance separation.)
For $\Phi(r)=b(r)=0$, our metric reduces to a flat deSitter metric;
while for $\chi=0$, it becomes the static wormhole metric Eq.
(2.1). We may let $\Phi(r)\rightarrow 0,b/r\rightarrow 0$ as
$r\rightarrow\infty$, so that the spacetime is asymptotically
deSitter or we may choose to let $\Phi(r),\,b(r)$ go to zero at
some finite value of $r$, outside of which the metric is deSitter.
The latter (together with a few other conditions) would correspond
to a cutoff of the wormhole material at some fixed radius.
Examples of each of these choices are given by Eqs. (2.16-2.18) and
Eqs. (2.19-2.22), respectively. However, our scheme should work for
any of the original MT metrics. As before, we also
demand that $\Phi(r)$ be everywhere finite, so that the only
horizons present are cosmological. The spacetime described by
Eq. (2.23), unlike the usual flat deSitter spacetime,
is inhomogeneous due to the presence of the wormhole.

      Our primary goal in this investigation is to use inflation to
enlarge an initially small (possibly submicroscopic) wormhole. We
choose $\Phi(r)$ and $b(r)$ to give a reasonable wormhole at $t=0$,
which we assume to be the onset of inflation.  To see that the
wormhole expands in size, consider the proper circumference $c$ of
the wormhole throat, $r=b=b_o,$ for $\theta=\pi/2$, at any time
$t=const.$:
    $$c=\int_0^{2\pi} e^{\chi t}\, b_o\,d\phi
       =e^{\chi t}\,(2\pi b_o)\,.             \eqno(2.24)$$
This is simply $e^{\chi t}$ times the initial circumference.
The radial proper length through the wormhole between any two
pts. $A$ and $B$ at any $t=const.$ is similarly given by:
 $$l(t)=\pm e^{\chi t} \int_{r_A}^{r_B}
               {dr\over{(1-b(r)/r)^{1/2}}}\,, \eqno(2.25)$$
which is just $e^{\chi t}$ times the initial radial proper
separation. Thus we see that both the size of the throat and the
radial proper distance between the wormhole mouths increase
exponentially with time.

        To see that the ``wormhole'' form of the metric is
preserved with time, let us embed a $t=const.,\,\theta=\pi/2$
slice of the spacetime given by Eq. (2.23) in a flat 3D Euclidean
space with metric:
        $$ds^2=d{\bar{z}}^2+d{\bar{r}}^2+{\bar{r}}^2\,{d\phi}^2\,.
                                                      \eqno(2.26)$$
The metric on our slice is:
        $$ds^2={e^{2\chi t}\,{dr^2}\over{(1-b(r)/r)}} +
                       e^{2\chi t}\, r^2\,{d\phi}^2\,.\eqno(2.27)$$
Comparing the coefficients of ${d\phi}^2$, we have
        $$\bar{r}={e^{\chi t}\,r}|_{t=const.}       \eqno(2.28)$$
    $${d\bar{r}}^2=e^{2\chi t}\,{dr}^2|_{t=const.} \eqno(2.29)$$
With respect to the ${\bar{z}},{\bar{r}},\phi$ coordinates, the
``wormhole'' form of the metric will be preserved if the metric
on the embedded slice has the form:
 $$ds^2={{d{\bar{r}}^2}\over{(1-{\bar{b}(\bar{r})/{\bar{r}})}}} +
                      {\bar{r}}^2{d\phi}^2\,,      \eqno(2.30)$$
where $\bar{b}(\bar{r})$ has a minimum at some
$\bar{b}(\bar{r}_o)=\bar{b}_o=\bar{r}_o$. We can rewrite Eq. (2.27)
in the form Eq. (2.30) by using Eqs. (2.28-9) and
         $$\bar{b}(\bar{r})=e^{\chi t}\,b(r).   \eqno(2.31)$$
In particular, one can easily show that Eq. (2.31) is satisfied
for the specific choices of $b(r)$ given by Eqs. (2.16) and (2.19)
by rewriting the right-hand sides of these equations in terms of
$\bar{r}$ and using Eq. (2.28).
        The inflated wormhole will have the same overall size and
shape {\it relative to the ${\bar{z}},{\bar{r}},\phi$ coordinate
system}, as the initial wormhole had relative to the initial
$z,r,\phi$ embedding space coordinate system. This is because
the embedding scheme we have presented corresponds to an embedding
space (or more properly, a series of embedding spaces, each
corresponding to a particular value of $t=const.$) whose $z,r$
coordinates ``scale'' with time. To see this, we can follow the
embedding procedure outlined in Eqs. (2.4-2.6), but using Eqs.
(2.26) and (2.30). It is readily apparent that
$${{d{\bar{z}}}\over{d{\bar{r}}}}
=\pm\left({{\bar{r}}\over{\bar{b}(\bar{r})}}-1\right)^{-1/2}
={{dz}\over{dr}} \,,                             \eqno(2.32)$$
where we have used Eqs. (2.28), (2.29), and (2.31). Eq. (2.32)
implies
$$\bar{z}(\bar{r})=\pm\int{{d\bar{r}}\over{(\bar{r}/{\bar{b}(\bar
{r})}-1)^{1/2}}}=
\pm e^{\chi t}\,\int{{dr}\over{(r/{b(r)}-1)^{1/2}}}
=\pm e^{\chi t}\,z(r)\,.                      \eqno(2.33)$$
Therefore, we see that the relation between our embedding space
at any time $t$ and the initial embedding space at $t=0$ is, from
Eqs. (2.29) and (2.33):
  $$ds^2=d{\bar{z}}^2+d{\bar{r}}^2+{\bar{r}}^2\,{d\phi}^2
        =e^{2\chi t}\,[dz^2+dr^2+r^2{d\phi}^2]\,.  \eqno(2.34)$$

  It is quite important to keep in mind (especially when
taking derivatives) that {\it Eqs. (2.28-9) do
not represent a ``coordinate transformation'', but rather a
``rescaling'' of the $r$-coordinate on each $t=const.$ slice}.
Relative to the ${\bar{z}},{\bar{r}},\phi$ coordinate system
the wormhole will always remain the same size; the scaling of
the embedding space compensates for the expansion of the wormhole.
Of course, the wormhole will change size relative to the initial
$t=0$ embedding space.

     If we write the analog of the ``flareout condition'', Eq.
(2.7), for the expanded wormhole we have
$${{d\,^2{\bar{r}(\bar{z})}}\over{d{\bar{z}}^2}}>0\,, \eqno(2.35)$$
at or near the throat. From Eqs. (2.28), (2.29), (2.31), and (2.32)
it follows that
     $${{d\,^2{\bar{r}(\bar{z})}}\over{d{\bar{z}}^2}}
       =e^{-\chi t}\,\left({{b-b'r}\over{2b^2}}\right)
       =e^{-\chi t}\,\left({{d\,^2r(z)}\over{dz^2}}\right)>0\,,
                                                      \eqno(2.36)$$
at or near the throat. Rewriting the right-hand side of Eq. (2.36)
relative to the barred coordinates, we obtain
     $${{d\,^2{\bar{r}(\bar{z})}}\over{d{\bar{z}}^2}}
=\left({{\bar{b}-{\bar{b}}'\bar{r}}\over{2{\bar{b}}^2}}\right)>0\,,
                                                     \eqno(2.37)$$
at or near the throat, where we have used Eqs. (2.28), (2.31), and
      $${\bar{b}}'(\bar{r})={{d\bar{b}}\over{d\bar{r}}}
          =b'(r)={{db}\over{dr}}\,.               \eqno(2.38)$$
We observe that relative to the barred coordinates, the ``flareout
condition'' Eq. (2.37), has the same form as that for the static
wormhole. With respect to the unbarred coordinates, the flareout
condition Eq. (2.36), appears as though it might be harder to
satisfy as time goes on because of the decaying exponential factor.
However, this is due to the fact that as the wormhole inflates,
its throat size and proper length inflate along with the
surrounding space. It therefore necessarily needs to ``flare
outward'' less and less at its throat as the two external spaces
connected by the wormhole move farther apart (again, relative
to the initial ``$t=0$'' embedding space). This behavior is
confirmed in an animated ``toy'' model of an inflating
wormhole produced with Mathematica [30], where $b(r)$ is given by
Eq. (2.16) [31].

        Let us now examine the stress-energy tensor that gives
rise to the wormhole described by Eq. (2.23). First, switch to a
set of orthonormal basis vectors defined by
    $$\eqalign{e_{\hat{t}}&=e^{-\Phi}\,e_{t},\cr
    e_{\hat{r}}&=e^{-\chi t}\,(1-b/r)^{1/2}\,e_{r},\cr
    e_{\hat{\theta}}&=e^{-\chi t}\,r^{-1}\,e_{\theta},\cr
    e_{\hat{\phi}}&=e^{-\chi t}\,(r\,sin\theta)^{-1}\,e_{\phi}.\cr}
                                                   \eqno(2.39)$$
This basis represents the proper reference frame of a set of
observers who always remain at rest at constant
$r,\,\theta,\,\phi$. The Einstein field equations will be written
in the form
    $$G_{\hat{\mu}\hat{\nu}}=R_{\hat{\mu}\hat{\nu}}
          -{1\over2}g_{\hat{\mu}\hat{\nu}}R
          ={8\pi}T_{\hat{\mu}\hat{\nu}}\,,         \eqno(2.40)$$
so that any ``cosmological constant'' terms will be incorporated
as part of the stress-energy tensor $T_{\hat{\mu}\hat{\nu}}$.
The components of $T_{\hat{\mu}\hat{\nu}}$ are
 $$\eqalignno{T_{\hat t \hat t}&=\rho(r,t)
 ={1\over{8\pi}}\left[3{\chi}^2\,e^{-2\Phi}
   +\,e^{-2\chi t}\,\,{b'\over{r^2}}\right] &(2.41) \cr
\noalign{\smallskip}
T_{\hat r \hat r}&=-\tau(r,t)
 ={1\over{8\pi}}\left[-3{\chi}^2\,e^{-2\Phi}
 -e^{-2\chi t}\,\left[{b\over{r^3}}-{{2{\Phi}'}\over r}\,
 \left(1-{b\over r}\right)\right]\right]    &(2.42)   \cr
\noalign{\smallskip}
 T_{\hat t \hat r}&=-f(r,t)
 ={1\over{8\pi}}\left[2e^{-\Phi-{\chi t}}\, \left(1-{b\over
r}\right)^{1/2}\,{\chi}\,{\Phi}'\right] &(2.43)\cr
\noalign{\smallskip}
 T_{\hat \theta \hat \theta}&=T_{\hat \phi \hat \phi}
 =p(r,t) \cr
\noalign{\smallskip}
&={1\over{8\pi}}\biggl[-3{\chi}^2\,e^{-2\Phi}\,+
\,e^{-2\chi t}\,\biggl[{1\over 2}\left({b\over{r^3}}-
{{b'}\over{r^2}}\right)
+{{{\Phi}'}\over r}\,\left(1-{b\over2r}-
{b'\over 2}\right) \cr
\noalign{\smallskip}
&\qquad+\left(1-{b\over r}\right)
[{\Phi}''+({\Phi}')^2] \biggr]\biggr].      &(2.44) \cr} $$
The quantities $\rho,\,\tau\,,f$, and $p$ are respectively: the
mass-energy density, radial tension per unit area, energy flux in
the (outward) radial direction, and lateral pressures as measured
by observers stationed at constant $r,\,\theta,\,\phi$.
Note from Eq. (2.43) that the flux vanishes at the wormhole throat,
as it must by symmetry. If we let $\Phi(r)\rightarrow
0,\,b/r\rightarrow 0$ as $r\rightarrow\nobreak\infty$, \hfil\break
the stress-energy
tensor components asymptotically assume their deSitter forms, i.e.,
\hfil\break $T_{\hat t\hat t}=-T_{\hat r\hat r}=-T_{\hat \theta\hat
\theta}=-T_{\hat \phi\hat \phi}=3{\chi}^2$.
Alternatively, we may wish to cutoff the wormhole material at some
fixed radius, $r=R$.
A sufficient condition for doing this would be to let
$\Phi(r)\,={\Phi}'\,={\Phi}''\,=\,b=\,b'=\,0$ for $r\geq R$.
For completeness, the Riemann curvature tensor components are
also included in an appendix. Note that all the stress-energy and
curvature components are finite for all $t$ and $r$.
For $\chi=0$, our expressions reduce to those of MT [1]. (Note the
correction of a sign error in the $({{\Phi}'\,b}/{2r^2})$ term of
$G_{\hat{\theta}\hat{\theta}}$ in their \hfil\break Eq. (12).)

\vskip 0.5in
\centerline{C. Simple Examples: the $\Phi(r)=0$ Cases}
\vskip 0.2in
      A particularly simple example of an inflating wormhole
is obtained by setting $\Phi(r)=0$ in Eq. (2.23):
 $$ds^2=-dt^2+e^{2\chi t}\left[{{dr^2}\over{(1-b(r)/r)}}
   +r^2({d\theta}^2+ sin^2\theta\,{d\phi}^2)\right]\,.
                                                   \eqno(2.45)$$
 The stress-energy tensor
components in an orthonormal frame (Eq. (2.39) with $\Phi=0$)
become
 $$\eqalignno{T_{\hat t \hat t}&=\rho(r,t)
 ={1\over{8\pi}}\left[3{\chi}^2\,
   +\,e^{-2\chi t}\,\,{b'\over{r^2}}\right] &(2.46) \cr
\noalign{\smallskip}
T_{\hat r \hat r}&=-\tau(r,t)
 ={1\over{8\pi}}\left[-3{\chi}^2\,
 -e^{-2\chi t}\,\left({b\over{r^3}}\right)\right]    &(2.47)   \cr
\noalign{\smallskip}
 T_{\hat t \hat r}&=-f(r,t)=0   &(2.48)   \cr
\noalign{\smallskip}
 T_{\hat \theta \hat \theta}&=T_{\hat \phi \hat \phi}
 =p(r,t)
={1\over{8\pi}}\biggl[-3{\chi}^2\,+
\,{{e^{-2\chi t}}\over 2}\left({b\over{r^3}}-
{{b'}\over{r^2}}\right)\biggr].      &(2.49) \cr} $$
The Riemann curvature tensor components for this metric are
also included in the appendix. Note that the stress-energy tensor
and Riemann tensor components all approach their deSitter space
values for large $t$. (The same is true for the expressions of
these quantities associated with the metric Eq. (2.23), modulo
some multiplicative factors of $e^{-\Phi}$, which would go to $1$
outside the ``wormhole'' part of the spacetime, e.g., at large
$r$.) When $\chi=0$, our metric reduces to that of a static
``zero-tidal-force'' wormhole, Eq. (2.12).
\eject

\centerline{\bf III. PROPERTIES OF THE SOLUTIONS AND DISCUSSION}
\vskip 0.2in
     A noticeable difference between the stress-energy tensors
associated with the $\Phi(r)\neq 0$ versus the $\Phi=0$ wormholes
is the presence of a flux term, given by Eq. (2.43). To understand
this, we must clarify the difference between two ``natural''
coordinate systems associated with the wormhole. The first can be
thought of as the rest frame of the wormhole geometry, i.e., an
observer at rest in this frame is at constant $r,\,\theta,\,\phi$.
The second can be thought of as the rest frame of the wormhole
material. In the absence of a particulate model for the wormhole
material, the best we can do is to define such a rest frame in
terms of the properties of the stress-energy tensor. More
specifically, we can define the rest frame of the wormhole material
as the one in which an observer co-moving with the material sees
zero energy flux.  From Eq. (2.43) we see that for $\Phi(r)\neq 0$,
the wormhole material is not at rest in the $r,\,\theta,\,\phi$
coordinate system. For the $\Phi(r)=0$ metrics given by Eq. (2.45),
the two coordinate systems coincide.

     Let $U^{\mu}=dx^{\mu}/{d\tau}=(U^{\,t},0,0,0)=
(e^{-\Phi(r)},0,0,0)$ be the four-velocity of an observer who is
at rest with respect to the $r,\,\theta,\,\phi$
coordinate system. The observer's four-acceleration is
$$\eqalignno{a^{\mu}&={{DU^{\mu}}\over {D\tau}}  \cr
\noalign{\smallskip}
         &=U^{\mu}\,_{;\,\nu}\,U^{\nu}          \cr
\noalign{\smallskip}
&=(U^{\mu}\,_{,\,\nu}+\Gamma^{\mu}_{\beta\nu}\,U^{\beta})\,U^{\nu}
                                       \,, &(3.1)   \cr}$$
which for the metric Eq. (2.23) gives the components
   $$\eqalignno{a^t&=0  \cr
\noalign{\smallskip}
 a^r&=\Gamma^r_{tt}\,\left({dt\over{d\tau}}\right)^2   \cr
\noalign{\smallskip}
          &=e^{-2\chi t}\,{\Phi}'\,(1-b/r)\,.        &(3.2) \cr}$$
{}From the geodesic equation, a radially moving test particle which
is initially at rest has the equation of motion
 $${{d^{\,2}r}\over d{\tau}^2}=-\Gamma^r_{tt}\,
\left({dt\over{d\tau}}\right)^2 =-a^r\,.
\eqno(3.3)$$
Therefore, we see that $a^r$ is the radial component of proper
acceleration that an observer must maintain in order to remain at
rest at constant $r,\,\theta,\,\phi$. From Eq. (3.3) it follows
that for $\Phi(r)\neq 0$ wormholes
(whether static or inflating), such observers do not
move geodesically (except at the throat), whereas for $\Phi(r)=0$
wormholes, they do. In the $\Phi(r)\neq 0$ case, for
observers at fixed $r,\,\theta,\,\phi$:
    $${\partial\over\partial r}\left(
     {{d\tau}\over dt}\right)={\Phi}'\,e^{\Phi(r)}\,. \eqno(3.4)$$
Eq. (3.4) can be thought of as the ``radial gradient of the flow
of proper time with respect to coordinate time''. Note that the
flux component of the stress-energy tensor, Eq. (2.43), goes like
$\chi\,{\Phi}'$. It therefore depends both on the time-dependence
of the spatial part of the metric and on the ``radial gradient
of proper time flow''.

      A wormhole will be called ``attractive'' if $a^r>0$
(observers must maintain an outward-directed radial acceleration
to keep from being pulled into the wormhole), and ``repulsive''
if $a^r<0$ (observers must maintain an inward-directed radial
acceleration to avoid being pushed away from the wormhole).
For $a^r=0$, the wormhole is neither attractive nor repulsive.
The sign of the energy flux depends on the sign of ${\Phi}'$, or
equivalently on the sign of $a^r$. Since the flux
$f=-T_{\hat t \hat r}$, then from Eq. (2.43) we see that if the
wormhole is attractive, there is a negative energy flow out
of it (or equivalently, a positive energy flow into it); if it is
repulsive, there is a negative energy flow into it (positive
energy flow out of it). In the case where the wormhole material is
cut off at a finite radius $r=R$, the energy flux vanishes at
both $r=R$ and $r=b=b_o$, though not necessarily in between.
For this situation, we might think of the flux as being due to
a redistribution of energy within the wormhole caused by its
expansion.

      The exoticity function, Eq. (2.9), of MT can be written:
   $$\zeta={{-T_{\hat{\mu}\hat{\nu}}\,W^{\hat{\mu}}
\,W^{\hat{\nu}}}\over{|T_{\hat t\hat t}|}}\,,
\eqno(3.5)$$
where $W^{\hat{\mu}}=(W^{\,\hat t},W^{\,\hat r},0,0)=(1,\pm 1,0,0)$
is a radial outgoing (ingoing) null vector. This condition is,
in some sense, a measure of the degree to which the wormhole
material violates the WEC. In our case,
$$\zeta={(\tau-\rho\mp f)\over {|\rho|}}\,.     \eqno(3.6)$$
{}From Eqs. (2.41-3), it can be shown that
$$\eqalignno{T_{\hat{\mu}\hat{\nu}}\,W^{\hat{\mu}}\,W^{\hat{\nu}}&=
{e^{-2\chi t}\over 8\pi}
\biggl[\biggl({b'\over{r^2}}-{b\over{r^3}}\biggr)
-{2{\Phi}'\over r}\,\biggl(1-{b\over r}\biggr)\biggr]  \cr
\noalign{\smallskip}
&\qquad\pm{e^{-\chi t}\over 4\pi}
\biggl[\biggl(1-{b\over r}\biggr)^{1/2}\,
\chi\,{\Phi}'\,e^{-\Phi}\biggr]\,.            &(3.7)  \cr}$$
For $\Phi(r)=0$, Eq. (3.6) reduces to
$$T_{\hat{\mu}\hat{\nu}}\,W^{\hat{\mu}}\,W^{\hat{\nu}}
={e^{-2\chi t}\over 8\pi}
\biggl({b'\over{r^2}}-{b\over{r^3}}\biggr)\,.       \eqno(3.8)$$
Using Eq. (3.6), (3.7), and (2.8), the exoticity function at any
radius and time can be written as
$$\eqalignno{\zeta &={{e^{-2\chi t}
\biggl[(2b^2/{r^3})
({d\,^2r(z)}/{dz^2})
+(2{\Phi}'/r)\,(1-b/r)\biggr]}\over
{\biggl|3{\chi}^2\,e^{-2\Phi}
+\,e^{-2\chi t}\,\,(b'/{r^2})\biggr|}} \cr
\noalign{\smallskip}
&\qquad \mp {{2e^{-{\chi t}}\,\biggl[(1-b/r)^{1/2}\,
{\chi}\,{\Phi}'\,e^{-\Phi}\biggr]}\over
{\biggl|3{\chi}^2\,e^{-2\Phi} +
\,e^{-2\chi t}\,\,(b'/{r^2})\biggr|}} \,.    &(3.9) \cr}$$

  Comparing Eq. (2.10) with Eq. (3.9), we see that the
relationship between the exoticity function and the flareout
condition does not seem to be quite as simple as that for the
static wormhole. The interpretation of Eq. (3.9) is complicated by
the presence of the ${\chi}^2$ term in the denominator, which could
have the opposite sign from the $b'$ term when the sign of the
latter is negative, as well as by the addition of the flux term.
If $3{\chi}^2\,e^{-2\Phi}\,\neq \,e^{-2\chi t}\,(b'/{r^2})$ for
all $t$, then from Eq. (2.41), $\rho$ is non-zero and finite. In
this case, the
vanishing of terms such as ${\Phi}'\,(1-b/r)$ at the throat
and Eq. (2.8) allow us to write that
     $$\zeta_o>0\,\,\,\, {\rm at\,or\,near\,the\,throat}\,,
                                     r=b=b_o\,.  \eqno(3.10)$$
If $\rho$ is non-zero and finite for all $t$, then
it can be shown from  Eq. (3.9) that the exoticity at the throat
$\zeta_o$, decays exponentially at large $t$. This is not terribly
surprising in light of our earlier discussion regarding the
``flareout'' behavior of the wormhole throat during inflation.

  Rather than examining the exoticity function, it is much simpler
to just look at the WEC along the null vectors $W^{\hat{\mu}}$ in
the limit $r\rightarrow b_o$. At the throat this condition,
$T_{\hat{\mu}\hat{\nu}}\,W^{\hat{\mu}}\,W^{\hat{\nu}}\geq\nobreak
0$,
simply reduces to the right-hand side of Eq. (3.8) evaluated at
$r=b=b_o$, for both the $\Phi\neq 0$ and $\Phi=0$ cases. The term
in parentheses is just the value of this expression at $t=0$, which
is the same as that for the static wormhole and thus
must be negative, from the original argument of MT. Therefore,
the violation of the WEC at the throat of the wormhole decreases
exponentially with time.

    To understand this behavior, one can give the following
heuristic argument. Consider the simple static $\Phi=0$ wormhole
example given by Eqs. (2.13-2.16), for different throat sizes. For
such a wormhole, the
negative energy density, radial tension per unit area, and lateral
pressure at the throat scale like $1/{b_o}^2$. They decrease in
magnitude as the size of the throat increases. (Note however,
that for this wormhole the exoticity $\zeta_o$ is independent of
throat size.) This makes sense because the smaller the wormhole
throat, the smaller its radius of curvature and hence the larger
the curvature. The larger the curvature, the more ``prone'' is the
wormhole to gravitational collapse, and therefore the larger
the negative energy density required to hold it open. However,
the total amount of negative energy near the throat scales like
$\rho V\sim  (1/{b_o}^2\,\times\,{b_o}^3) \sim b_o$, and therefore
must increase as the throat size increases.

  In general, due to the rapid expansion of the surrounding space,
the two mouths of the wormhole will quickly lose causal contact
with one another, i.e., they will move outside of each other's
cosmological horizon. Each mouth might re-enter the other's horizon
after inflation [32]. If the mouths were to remain in causal
contact throughout the duration of the inflationary period, then
there would be a constraint on the initial size of the wormhole. To
estimate this, we will use the simple $\Phi(r)=0$ wormhole metric,
Eq. (2.45). Consider two observers stationed on opposite sides of
the wormhole and separated by an initial radial proper distance at
$t=0$ of $l_o$. Let $l(T)$ be their separation at the end of
inflation, $t=T$. The proper distance, $l_H$, of each observer from
his/her horizon is $l_H\sim {1/{\chi}}$. If we require that this
distance be less than $l(T)$, then
        $$l_o < {e^{-\chi T}\over{\chi}}\,.       \eqno(3.11)$$
For a typical inflationary scenario (see for example, [33]),
${\chi}^{-1}\sim 10^{-34}\,{\rm sec}\sim\nobreak
10^{-23}\,{\rm cm}\,,\hfil\break
{\chi T}\sim 100\,$, which gives $l_o < 10^{-67}\,{\rm cm}\,<<l_P
\sim  10^{-33}\,{\rm cm}\,$. Since the Planck length, $l_P$, is
usually regarded as the smallest distance scale which makes
physical sense, it seems that the condition Eq. (3.11) cannot
be satisfied (at least in the usual inflationary scenarios).
The same parameters yield an increase in size of the wormhole
by a factor of $\sim 10^{43}$. A initially Planck-sized wormhole
would be enlarged to a size of $\sim 10^{10}\,{\rm cm}\,\sim 1\,
R_{\odot}$ after inflation.

   It is also possible that the wormhole
will continue to be enlarged by the subsequent FRW phase of
expansion. One could perform a similar analysis to ours by
replacing the deSitter scale factor in Eq. (2.23) by an
FRW scale factor $a(t)$.
A naive estimate yields a total enlargement of wormhole size which
is larger than our present horizon size. However, since it is
difficult to even say what effects the reheating at the end of
inflation will have on the wormhole, we will not pursue this
possibility further.

     Since the two mouths of the wormhole lose causal contact
during inflation, then presumably issues of traversability will
arise only after inflation. In our discussion we have therefore
avoided the enforcement of additional ``usability criteria'', i.e.,
requirements proposed by MT which are designed to make wormhole
traversal comfortable for human travellers. Also, a wormhole need
not necessarily be traversable by human beings for it to be useful.
Indeed, the more troubling characteristics of wormholes, such as
their use for possible causality violation, should be realizable
if it is possible to just send signals through them, in the form
of light rays or particles. In passing, we again note that the
Riemann curvature tensor components, given in the
appendix, are well-behaved for all $r$ and $t$ (e.g., no
``exponentially growing'' tidal forces at the throat).

    One might think that since two-way passage is practical only
after inflation, the application of the present scenario to
small ordinary Schwarzschild or Reissner-Nordstrom wormholes might
yield large wormholes which could then later be made traversable.
However, these wormholes have (non-cosmological) horizons which
tend to make them collapse very rapidly- an affliction which
would probably be exacerbated by the positive energy
released during the decay of the false vacuum. Assuming that
one could circumvent the latter problem, then perhaps such
a wormhole might be stabilized by the injection of a flux of
negative energy. Unfortunately, the magnitude and duration of such
fluxes would most likely be limited by ``quantum-inequality'' type
restrictions similar to those found to hold for negative fluxes
injected into an extreme Reissner-Nordstrom black hole [13]. The
same would likely be true for the pair-produced extreme
magnetically charged wormholes conjectured by Garfinkle and
Strominger [34].

     A nontrivial problem is the maintenance of the wormhole during
and after the decay of the false vacuum. We saw earlier that
although large (static) wormholes with $\rho<0$
required a smaller negative energy density for maintenance
than small ones, the total amount of negative energy required
should increase with increasing throat size. During inflation the
wormhole throat is greatly stretched in size due to the rapid
cosmological expansion. However, note that in Eqs. (2.41)-(2.44),
the ``false vacuum terms'' remain constant with time while the
``exotic wormhole material terms'' decay exponentially with time.
For example, in Eq. (2.41) the first term, which represents the
energy density of the false vacuum, remains constant (at constant
$r$) while the second term, representing the ``exotic'' energy
density of the wormhole, decreases with time. Consider the case
where the latter is negative. Then the total amount of positive
energy increases, since the positive energy density of the false
vacuum remains constant as the volume increases. The total amount
of negative energy decreases because the negative energy density
exponentially decreases while the volume increases.
When the false vacuum decays, the exponential stretching will cease
and the positive energy in the false vacuum will be converted into
more conventional forms, such radiation and/or particles. This
potentially huge positive energy might flood the wormhole,
triggering a gravitational collapse of the throat. Perhaps such
a fate might be avoided if the two energy densities in Eq. (2.41)
are roughly comparable in magnitude at the end of inflation.

       As a simple example, let us first consider the inflating
``absurdly benign'' wormhole with $\Phi$ and $b$ given by Eq.
(2.19). From Eq. (2.46), the energy density at the throat is
$$\rho_o\,\sim\,3{\chi}^2\,-\,2e^{-2\chi t}\,{(b_o\,a_o)}^{-1}\,,
                                                   \eqno(3.12)$$
where $a_o$ is the thickness (in $r$) of the negative energy
region near the throat. Let $a_o=\eta\,b_o$, where $\eta$ is
some fraction, but require $b_o<l_P$, $a_o<l_P$. For the two
terms in Eq. (3.12) to be comparable
at the end of inflation, $t=T$ :
$$b_o\,\sim\,{e^{-\chi T}\over{\eta\,\chi}}\,.       \eqno(3.13)$$
which is almost identical to the condition Eq. (3.11). For the
inflation parameters given earlier, we see that Eq. (3.13) also
leads to a required initial wormhole size $b_o<<\,l_P$.
One fares a little better with the $\Phi(r)\neq 0$ wormhole.
{}From Eq. (2.41), it appears that by making $\Phi(b_o)$
large enough it might be possible to suppress the positive
${\chi}^2$ ``false vacuum'' term. The energy density at the throat
goes like [35]:
$$\rho_o\,\sim\,3{\chi}^2\,e^{-2\Phi(b_o)}
   -\,({e^{-2\chi t}}\,/\,{b_o}^2)\,,             \eqno(3.14)$$
which leads to the condition that
$$b_o\,\sim\,e^{\Phi(b_o)}\,{e^{-\chi T}\over{\chi}}\,.
                                                 \eqno(3.15)$$
For $b_o\sim\,10^{-33}\,{\rm cm}\,$,
$\Phi(b_o)\,\sim\,{\rm ln}(10^{34})\,\sim 78$. This corresponds to
a time dilation factor of $(d\tau/dt)\,\sim\,10^{34}$, i.e., clocks
fixed at $r=b_o$ must run $\sim 10^{34}\,\times$ {\it faster}
than clocks outside the wormhole!

These crude heuristic arguments
suggest that in general it will be difficult for the negative
energy density-type terms to overwhelm the false vacuum-type terms.
However it should be mentioned that our simple argument does not
take into account the effects of gravitational energy, so it is
not completely clear as to whether wormholes are unlikely to
survive inflation. Also, the results in this paper represent only
one possible generalization of MT wormholes to time-dependent
situations. Even more general solutions might be obtained
by allowing $\Phi$ and $b$ in our metrics to be functions
of $t$ as well as $r$ [36].

      On the other hand, if most of the wormholes in the quantum
foam survived inflation, then the universe might be far more
inhomogeneous and topologically complicated than we observe [37]
(unless they all inflated beyond our current horizon).
Perhaps the wormholes were all destroyed by the flood of positive
energy released during reheating.
It is also possible that a given wormhole mouth might
find itself in a slightly different gravitational potential from
its counterpart. The quantum field-theoretic instabilities
associated with the tendency of such a wormhole to form closed
timelike curves [6,7,8] might destroy it.
Perhaps the probability for the existence of a wormhole in the
quantum foam that has the right properties for inflation is
extremely low, or perhaps none of the foam inflates (after all,
galaxies in the FRW phase don't expand). Since we know very little
about the quantum foam (or whether it even exists at all!), these
are difficult questions to answer. (The possibility of artificially
enlarging a tiny wormhole by imbedding it in a false vacuum bubble
is currently under investigation.)

   Another part of the problem is that one does not know what
constitutes a ``generic'' wormhole. In classical general
relativity, the energy conditions determine the characteristics
of ``reasonable'' sources. Quantum field theory allows {\it some}
violation of the energy conditions, but with our present state of
knowledge regarding the extent of these violations, we cannot
yet say which types of wormholes, if any, are physically
reasonable.

\vskip 0.5in
{\bf Acknowledgements:}
The author is grateful to Larry Ford and Kip Thorne for several
long and detailed discussions of this problem, and for helpful
comments on the presentation. I would also like to
thank Matt Visser, Eanna Flanagan, Ulvi Yurtsever,
David Garfinkle, Mike Morris, Ron Mallett, and Kristine Larsen
for valuable comments, and the Aspen Center for Physics, whose
hospitality made many of these discussions possible. This research
was supported in part by NSF Grant No. PHY-8905400 and
by an AAUP/CCSU Faculty Research Grant.
\vskip 0.5in
\eject
\centerline{\bf Appendix}
       The following curvature tensor components, as well as
some of the stress-energy tensor components found in the text,
were computed using MathTensor [38]. For the metric Eq. (2.23), the
Riemann tensor components are:
$$\eqalignno{R_{\hat{t}\hat{\phi}\hat{t}\hat{\phi}}
&=-R_{\hat{\phi}\hat{t}\hat{t}\hat{\phi}}
=-R_{\hat{t}\hat{\phi}\hat{\phi}\hat{t}}
=R_{\hat{\phi}\hat{t}\hat{\phi}\hat{t}}  \cr
\noalign{\smallskip}
&= -{\chi}^2\,e^{-2\Phi}\,+\,e^{-2\chi t}\,({\Phi}'/{r^2})\,
(r-b)                               &(A1) \cr
\noalign{\smallskip}
R_{\hat{t}\hat{\phi}\hat{\phi}\hat{r}}
&=R_{\hat{\phi}\hat{r}\hat{t}\hat{\phi}}
=-R_{\hat{\phi}\hat{t}\hat{\phi}\hat{r}}
=-R_{\hat{t}\hat{\phi}\hat{r}\hat{\phi}}
=R_{\hat{\phi}\hat{t}\hat{r}\hat{\phi}}    \cr
\noalign{\smallskip}
&= -\chi\,e^{-\chi t}\,(1-b/r)^{1/2}\,e^{-\Phi}\,{\Phi}' &(A2)\cr
\noalign{\smallskip}
R_{\hat{t}\hat{\theta}\hat{t}\hat{\theta}}
&=-R_{\hat{\theta}\hat{t}\hat{t}\hat{\theta}}
=-R_{\hat{t}\hat{\theta}\hat{\theta}\hat{t}}
=R_{\hat{\theta}\hat{t}\hat{\theta}\hat{t}}   \cr
\noalign{\smallskip}
&= -{\chi}^2\,e^{-2\Phi}\,+\,e^{-2\chi t}\,({\Phi}'/{r^2})\,
(r-b)                               &(A3) \cr
\noalign{\smallskip}
R_{\hat{t}\hat{\theta}\hat{\theta}\hat{r}}
&=R_{\hat{\theta}\hat{r}\hat{t}\hat{\theta}}
=-R_{\hat{\theta}\hat{t}\hat{\theta}\hat{r}}
=-R_{\hat{t}\hat{\theta}\hat{r}\hat{\theta}}
=R_{\hat{\theta}\hat{t}\hat{r}\hat{\theta}}      \cr
\noalign{\smallskip}
&= -\chi\,e^{-\chi t}\,(1-b/r)^{1/2}\,e^{-\Phi}\,{\Phi}' &(A4)\cr
\noalign{\smallskip}
R_{\hat{t}\hat{r}\hat{t}\hat{r}}
&=-R_{\hat{r}\hat{t}\hat{t}\hat{r}}
=-R_{\hat{t}\hat{r}\hat{r}\hat{t}}
=R_{\hat{r}\hat{t}\hat{r}\hat{t}}    \cr
\noalign{\smallskip}
&=-{\chi}^2\,e^{-2\Phi}\,+\,e^{-2\chi t}\,(1-b/r)\,[{\Phi}''
\,+\,({{\Phi}'}^2)]\,            \cr
\noalign{\smallskip}
&\qquad\quad+\,(e^{-2\chi t}/2)\,{\Phi}'\,(b/{r^2}-b'/r)
                                 &(A5) \cr
\noalign{\smallskip}
R_{\hat{\theta}\hat{r}\hat{\theta}\hat{r}}
&=-R_{\hat{r}\hat{\theta}\hat{\theta}\hat{r}}
=-R_{\hat{\theta}\hat{r}\hat{r}\hat{\theta}}
=R_{\hat{r}\hat{\theta}\hat{r}\hat{\theta}}  \cr
\noalign{\smallskip}
&= {\chi}^2\,e^{-2\Phi}\,+\,(e^{-2\chi t}/2)\,(b'/{r^2}-b/{r^3})
                                              &(A6)\cr
\noalign{\smallskip}
R_{\hat{\phi}\hat{\theta}\hat{\phi}\hat{\theta}}
&=-R_{\hat{\theta}\hat{\phi}\hat{\phi}\hat{\theta}}
=-R_{\hat{\phi}\hat{\theta}\hat{\theta}\hat{\phi}}
=R_{\hat{\theta}\hat{\phi}\hat{\theta}\hat{\phi}}  \cr
\noalign{\smallskip}
&= {\chi}^2\,e^{-2\Phi}\,+\,e^{-2\chi t}\,(b/{r^3}) &(A7) \cr
\noalign{\smallskip}
R_{\hat{\phi}\hat{r}\hat{\phi}\hat{r}}
&=-R_{\hat{r}\hat{\phi}\hat{\phi}\hat{r}}
=-R_{\hat{\phi}\hat{r}\hat{r}\hat{\phi}}
=R_{\hat{r}\hat{\phi}\hat{r}\hat{\phi}}  \cr
\noalign{\smallskip}
&= {\chi}^2\,e^{-2\Phi}\,+\,(e^{-2\chi t}/2)\,(b'/{r^2}-b/{r^3})\,.
                                     &(A8)  \cr}$$

  For the metric Eq. (2.45), the above components reduce to:
$$\eqalignno{R_{\hat{t}\hat{\phi}\hat{t}\hat{\phi}}
&=-{\chi}^2\,  &(A9)\cr
\noalign{\smallskip}
R_{\hat{t}\hat{\theta}\hat{t}\hat{\theta}}
&=-{\chi}^2\, &(A10) \cr
\noalign{\smallskip}
R_{\hat{t}\hat{r}\hat{t}\hat{r}}
&=-{\chi}^2\, &(A11) \cr
\noalign{\smallskip}
R_{\hat{\theta}\hat{r}\hat{\theta}\hat{r}}
&= {\chi}^2\,+\,(e^{-2\chi t}/2)\,(b'/{r^2}-b/{r^3})\, &(A12)\cr
\noalign{\smallskip}
R_{\hat{\phi}\hat{\theta}\hat{\phi}\hat{\theta}}
&= {\chi}^2\,+\,e^{-2\chi t}\,(b/{r^3})\, &(A13)\cr
\noalign{\smallskip}
R_{\hat{\phi}\hat{r}\hat{\phi}\hat{r}}
&= {\chi}^2\,+\,(e^{-2\chi t}/2)\,(b'/{r^2}-b/{r^3})\,.
&(A14)\cr}$$

\eject
\centerline{\bf References}
\vskip 0.3in
\item{[1]} M.S. Morris and K.S. Thorne, Am. J. Phys. {\bf 56}, 395
(1988).
\item{[2]} J. Friedman, M.S. Morris, I.D. Novikov, F. Echeverria,
G. Klinkhammer, K.S. Thorne, and U.Yurtsever, Phys. Rev. {\bf D42},
1915 (1990); M. Visser, Phys. Rev. {\bf D39}, 3182 (1989); {\bf
D41}, 1116 (1990); {\bf D43}, 402 (1991).
It appears as though a traversable wormhole metric may have been
first presented by H.G. Ellis, J. Math. Phys. {\bf 14}, 104 (1973).
See the letter by G. Clement, Am. J. Phys. {\bf 57}, 967 (1989),
and references listed there.

\item{[3]} S.W. Hawking and G.F.R. Ellis, {\it The Large Scale
Structure of Spacetime} (Cambridge University Press, London, 1973),
pp. 88-96.

\item{[4]} F.J. Tipler, Phys. Rev. {\bf D17}, 2521 (1978);
T. Roman, Phys. Rev. {\bf D33}, 3526 (1986); {\bf D37}, 546 (1988);
A. Borde, Class. Quantum Gravit. {\bf 4}, 343 (1987).

\item{[5]} M.S. Morris, K.S. Thorne, and U.Yurtsever, Phys. Rev.
Lett. {\bf 61}, 1446 (1988).

\item{[6]} V.P. Frolov and I.D. Novikov, Phys. Rev. {\bf D42}, 1057
(1990).

\item{[7]} S.W. Hawking, Phys. Rev. {\bf D46}, 603 (1992).
\item{[8]} S.W. Kim and K.S. Thorne, Phys. Rev. {\bf D43}, 3929
(1991); V.P. Frolov, Phys. Rev. {\bf D43}, 3878 (1991);
G. Klinkhammer, {\it ``Vacuum Polarization of Scalar and Spinor
Fields Near Closed Null Geodesics''} to be published in Phys.
Rev.D; J. Grant, Phys. Rev.D, submitted.

\item{[9]} H. Epstein, V. Glaser, and A. Jaffe, Nuovo Cim. {\bf
36}, 1016 (1965).

\item{[10]} H.B.G. Casimir, Proc. Kon. Nederl. Akad. Wetenschap
{\bf 51}, 793 (1948); L.S. Brown and G.J. Maclay, Phys. Rev.
{\bf 184}, 1272 (1969).
\item{[11]} R. Wald and U. Yurtsever, Phys. Rev. {\bf D44}, 403
(1991).
\item{[12]} L.H. Ford, Proc. Roy. Soc. Lond. {\bf A364}, 227
(1978); {\bf D43}, 3972 (1991).

\item{[13]} L.H. Ford and T. Roman, Phys. Rev. {\bf D41}, 3662
(1990); {\bf D46}, (1992).
\item{[14]} L.H. Ford and T. Roman, in preparation.
\item{[15]} R.P. Geroch, J. Math. Phys. {\bf 8}, 782 (1967).

\item{[16]} A. Borde, {\it ``Topology Change in General
Relativity''}, preprint.

\item{[17]} J.A. Wheeler, {\it Geometrodynamics} (Academic Press,
New York, 1962); J. Friedman, in {\it Conceptual Problems of
Quantum Gravity}, ed. A. Ashtekar and J. Stachel (Birkhauser,
Boston, 1991), pp. 540-572.
\item{[18]} S. Blau, E. Guendelman, and A. Guth, Phys. Rev.
{\bf D35}, 1747 (1987); E. Farhi and A. Guth, Phys. Lett.
{\bf B183}, 149 (1987).

\item{[19]} For a review, see S. Blau and A. Guth, in {\it 300
Years of Gravitation}, ed. S.W. Hawking and W. Israel (Cambridge
University Press, Cambridge, 1987), pp. 561-570.

\item{[20]} R. Basu, A. Guth, and A. Vilenkin, Phys. Rev. {\bf
D44}, 340 (1991).
\item{[21]} R. Mallett, Phys. Rev. {\bf D31}, 416 (1985);
{\bf D33}, 2201 (1986); see also K.Larsen and \hfil\break
R. Mallett, Phys. Rev. {\bf D44}, 333 (1991).

\item{[22]} K. Sato, M. Sasaki, H. Kodama, K. Maeda, Prog. Theor.
Phys. {\bf 65}, 1443 (1981);\hfil\break K. Maeda, K. Sato, M.
Sasaki, H. Kodama, Phys. Lett. {\bf B108}, 98 (1982); K.
Sato,\hfil\break
H. Kodama, M. Sasaki, K. Maeda, Phys. Lett. {\bf B108}, 103 (1982).

\item{[23]} S.W. Kim, Phys. Lett. {\bf A166}, 13 (1992).

\item{[24]} M. Visser, Nucl. Phys. {\bf B328}, 203 (1989).

\item{[25]} D. Hochberg, Phys. Lett. {\bf B251}, 349 (1990).

\item{[26]} D. Hochberg and T. Kephart, Phys. Lett. {\bf B268}, 377
(1991).

\item{[27]} C. Kuo and L.H. Ford, {\it ``Semiclassical
Gravity Theory and Quantum Fluctuations''}, manuscript in
preparation.

\item{[28]} C.W. Misner, K.S. Thorne, and J.A. Wheeler,
{\it Gravitation} (W.H. Freeman and Co., San Francisco, 1973)
pp. 612-615 and pp. 836-840.

\item{[29]} A similar technique was used by A. Bokhari and
A. Qadir, in {\it Proceedings of the Fourth Marcel Grossman
on General Relativity and Gravitation}, ed. R. Ruffini,
(Elsevier, 1986), pp. 1635-1642, to generalize the Schwarzschild
solution to an FRW-type background.

\item{[30]} Wolfram Research, Inc., {\it Mathematica}, (Wolfram
Research, Inc., Champaign, Illinois, 1991).

\item{[31]} T. Roman, to be published elsewhere.

\item{[32]} I am grateful to Kip Thorne for bringing this point
to my attention.

\item{[33]} E.W. Kolb and M.S. Turner, {\it The Early Universe}
(Addison-Wesley, New York, 1990), pp. 270-275.

\item{[34]} D. Garfinkle and A. Strominger, Phys. Lett. {\bf B256},
146 (1991).

\item{[35]} On dimensional grounds, $b'/{r^2}$ must $\sim
1/{b_o}^2$ at the throat. Here we have taken the case where $b'$ is
negative. However, it {\it need} not be. Recall that the violation
of the WEC at the throat involves the combination $\tau-\rho$ ($f$
vanishes at the throat). Violation of the WEC implies that {\it
some} observers will see negative energy densities, but those
observers need not be the ``co-moving'' ones.
\eject
\item{[36]} It is interesting that naively allowing $\Phi$ and $b$
to be functions of $t$ as well as $r$ in the original MT metric
does not give a well-behaved, time-dependent wormhole solution.
The flux term, $T_{\hat t \hat r}=(1/{8\pi})\,(b_{,t}/{r^2})\,
({\rm exp}[-2 \Phi(r,t)])\,(1-b/r)^{-1/2}$, diverges at\hfil\break
$r=b=b_o$ (T. Roman, unpublished). The problem seems to
come from the fact that if we demand that the geometric
significance of $r$ be that $2\pi r$ is the circumference of a
circle centered at the wormhole's throat, then the {\it only} such
circle which has any time-dependence is the one at the throat
$r=b_o(t)$.

\item{[37]} R. Penrose, in {\it Proceedings of the Fourteenth Texas
Symposium on Relativistic Astrophysics}, ed. E.J. Fenyves (Annals
of the New York Academy of Sciences, Vol. 571, New York, 1989),
pp. 262-3.

\item{[38]} L. Parker and S.M. Christensen, MathTensor
(MathSolutions, Inc., Chapel Hill, NC, 1992).

\bye